\documentclass[11pt]{article}

\usepackage[final]{acl}

\usepackage{times}
\usepackage{latexsym}

\usepackage[T1]{fontenc}

\usepackage[utf8]{inputenc}

\usepackage{microtype}

\usepackage{inconsolata}

\usepackage{graphicx}
\usepackage{amsmath}
\usepackage{amssymb}
\usepackage{mathtools}
\usepackage{amsthm}
\usepackage{makecell}
\usepackage{url}
\usepackage{multirow}
\usepackage{diagbox}
\usepackage{kotex}
\usepackage{xcolor}
\usepackage{verbatim}
\usepackage{listingsutf8}
\usepackage{beramono} 
\usepackage{listings}
\lstdefinestyle{prompt}{
  basicstyle=\ttfamily\small,
  keywordstyle=[1]\color{blue},
  breaklines=true,
  frame=single,
  literate={<}{{\textless}}1
           {>}{{\textgreater}}1
           {'}{'}1 {`}{`}1 {\_}{\textunderscore}1 {\{}{\{}{1} {\}}{\}}{1}
}

\usepackage[capitalize,noabbrev]{cleveref}

\title{FastSLM: Hierarchical Temporal Abstraction for Efficient Long-Form Speech Adaptation}

\author{
    \textbf{Junseok Lee}$^{1}$ \quad
    \textbf{Chang-Jae Chun}$^{2*}$ \\
    \\
    $^{1}$ OKESTRO Co., Ltd.\\
    $^{2}$ Department of Artificial Intelligence and Data Science, Sejong University\\
    {\footnotesize \texttt{js.lee6@okestro.com \quad\ cchun@sejong.ac.kr}
}
}

\begin{document}

\maketitle

\begingroup
\renewcommand\thefootnote{}
\footnotetext{* Corresponding author.}
\endgroup

\begin{abstract}
    Scaling Multimodal Large Language Models (MLLMs) to long-form speech is bottlenecked by the explosive growth of input tokens. Existing speech-language models project high-frame-rate acoustic features directly into the LLM input space, making long-context processing computationally prohibitive. Unlike images or videos, speech lacks spatial redundancy, making extreme token compression particularly challenging. To address this limitation, we propose FastSLM, a token-efficient architecture featuring the Hierarchical Temporal Abstractor (HTA), which progressively distills acoustic features across multiple temporal scales. HTA achieves an extreme compression rate of 1.67 tokens per second (97\% reduction) while preserving essential linguistic information for downstream speech-language understanding. Experimental results demonstrate that FastSLM achieves competitive performance across diverse speech-language tasks while requiring substantially fewer speech tokens and FLOPs than existing speech-language models. The source code and model checkpoints are available at \url{https://github.com/Lee-junseok1025/FastSLM}.
\end{abstract}

\section{Introduction}
\label{sec:intro}
Large Language Models (LLMs) \citep{gpt4,llama3,gemini,qwen3} have demonstrated remarkable reasoning capabilities, prompting research into Multimodal LLMs (MLLMs) for vision, audio, and video \citep{mllm_survey_1,mllm_survey_2}. Since speech is the primary interface for human-AI interaction \citep{MMAU}, adapting LLMs to the speech domain via Audio Language Models (ALMs) has become a central focus \citep{salmonn,phi4-m,qwen-audio,qwen2-audio,AF3}.

However, a fundamental challenge hinders ALM scalability: the granularity mismatch between acoustic features and textual tokens. Typical approaches project dense frame-level features from encoders like Whisper \citep{Whisper} directly into the LLM input space. For multi-minute audio, this sequence length explodes, rendering the quadratic computational cost of the autoregressive LLM prohibitive \citep{transformers}. 

A critical challenge in extreme speech compression stems from the inherent nature of the audio modality. Unlike images or videos with high spatial redundancy, speech is a sequential, non-overlapping stream of localized acoustic events. Attempting to compress this dense stream into a highly reduced token space using flat pooling or single-stage projection inevitably causes catastrophic loss of fine-grained acoustic cues (e.g., phonemes, intonation) necessary for linguistic understanding. 

To overcome this, we propose a shift from flat feature alignment to progressive temporal abstraction. We present FastSLM, a token-efficient architecture designed to distill long-form speech into highly compact representations. We introduce the Hierarchical Temporal Abstractor (HTA), which progressively compresses non-overlapping acoustic features from local details into global contexts. 

Crucially, this progressive mechanism facilitates robust Hierarchical Feature Compression. By structuring compression hierarchically, HTA progressively transitions from capturing fine-grained acoustic cues at early stages to extracting global semantic representations at the deeper stages. Through this, we achieve an effective compression rate of 1.67 tokens per second. This rate closely matches typical human speech rates for clear articulation \citep{speech_rate,speech_rate2}, effectively establishing a natural information bottleneck that synchronizes continuous audio streams with the discrete semantic processing pace of LLMs.

Our main contributions are summarized as follows:
\begin{itemize}
\setlength\itemsep{0em}
    \item \textbf{Hierarchical Temporal Abstraction (HTA):} We introduce the Hierarchical Temporal Abstractor (HTA), which reformulates speech--LLM alignment as progressive hierarchical abstraction rather than flat compression. HTA reduces frame-level speech representations by 97\% to 1.67 tokens/sec while preserving linguistic information across multiple temporal scales.  
    \item \textbf{Cost-Efficient Adaptation Strategy:} We demonstrate a three-stage training pipeline that leverages accessible automatic speech recognition (ASR) corpora for long-form speech adaptation, reducing the reliance on scarce and costly long-context instruction data.
    \item \textbf{Empirical Efficiency and Scalability:} FastSLM achieves competitive performance across multi-task benchmarks while exhibiting near-linear memory scaling, enabling efficient processing of hour-long audio under practical memory constraints.
\end{itemize}

\section{Related Work}

\subsection{Audio and Speech Language Models}
To advance LLMs toward comprehensive multimodal understanding, numerous studies have explored extending them to the audio modality. Models such as AudioPaLM~\citep{AudioPaLM}, Kimi-Audio~\citep{Kimi-audio}, and Qwen-Audio~\citep{qwen-audio,qwen2-audio} have demonstrated that integrating frame-level speech features into the LLM embedding space enables effective end-to-end spoken understanding. 

However, most of these models are primarily trained on short-form speech (typically under 60 seconds), severely limiting their ability to process multi-minute inputs. To address this limitation, recent models like Voxtral~\citep{Voxtral} and Audio-Flamingo3~\citep{AF3} introduced long-context training strategies and frame-level cross-attention for modality adaptation. Despite their strong progress, these approaches still rely on dense frame-level cross-attention or repeated window-level processing. This reliance introduces prohibitive computational overhead when handling multi-minute speech, failing to resolve the fundamental challenge of efficiently aligning long-form speech with LLMs under strict FLOPs constraints.

\subsection{Multi-modal Token Compression for Speech Modality Adaptation}

To address the token explosion problem in speech-language models, recent studies have introduced query-based compression modules that transform dense frame-level speech features into a compact set of learnable query tokens. SALMONN~\citep{salmonn} first adopted a window-level Q-Former for speech-language alignment, while subsequent segment-level Q-Former~\citep{seg-former} partitioned speech into fixed-length segments prior to compression.

More recent methods have further improved query-based compression from different perspectives. MMCE-QFormer~\citep{MMCE-QFormer} incorporates multimodal contextual information to jointly model speech and text, CompressedToFindLM~\citep{STP} introduces prototype-guided compression for local frame representations, and AlignFormer~\citep{Alignformer} combines CTC supervision with a dynamic-window Q-Former to improve temporal alignment for autoregressive decoding. Recent long-form speech models, including FastLongSpeech~\citep{fastlongspeech} and SpeechPrune~\citep{speechprune}, have also explored efficient long-form processing through token reduction and adaptive pruning strategies.

Despite these advances, most existing approaches perform token compression in a single stage over local windows or segments, which may limit aggressive compression for long-form speech understanding. In contrast, our HTA performs progressive multi-stage temporal abstraction across multiple temporal scales, enabling a 97\% token reduction while maintaining competitive performance across diverse speech-language tasks.

\section{Methodology}
In Section \ref{sec:model_archi}, we describe the architecture of FastSLM and the inference process. In Section \ref{sec:training}, we describe the training strategy employed for speech modality adaptation.

\subsection{Model Architecture of FastSLM}
\label{sec:model_archi}
\begin{figure}[!h]
    \centering
    \includegraphics[width=\columnwidth]{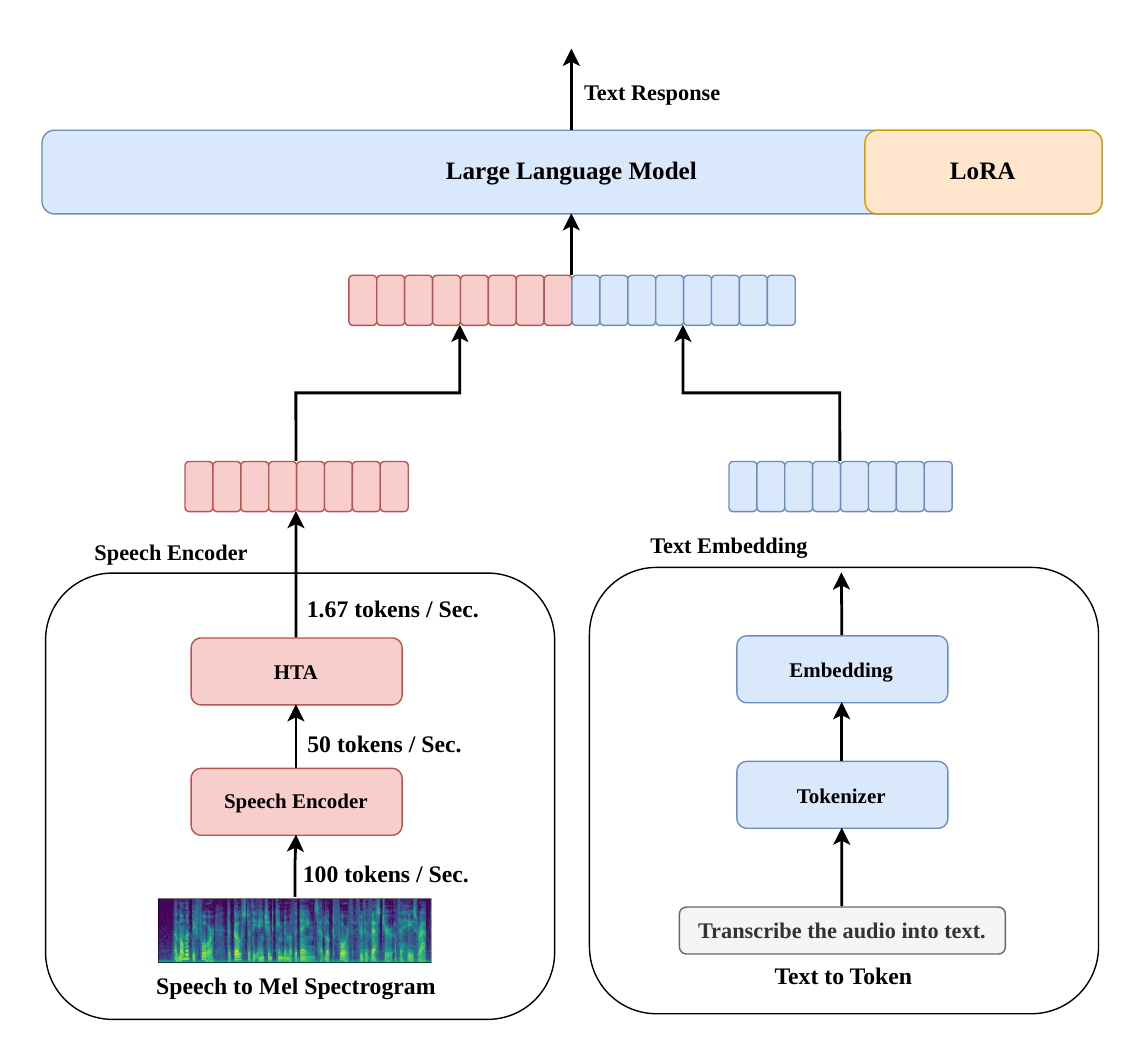}
    \caption{Architecture of FastSLM.}
    \label{fig:FastSLM}
\end{figure}

\textbf{FastSLM Overview}: The overall architecture of FastSLM is illustrated in Fig. \ref{fig:FastSLM}. Unlike conventional SLMs that directly concatenate dense acoustic features with text embeddings, FastSLM introduces a dedicated compression bottleneck to bridge the information density gap between modalities. The pipeline proceeds as follows: Raw waveforms are converted into Mel spectrograms (25-ms window and a 10-ms stride) and processed by a speech encoder \citep{Whisper} to extract high-fidelity frame-level features (50 Hz). Crucially, instead of feeding these dense frames to the LLM, we employ the HTA to distill them into highly compact semantic tokens. This mechanism achieves an extreme compression rate of approximately 1.67 tokens per second, reducing the sequence length by over 97\% compared to the original encoder output. These compressed tokens are then concatenated with text prompts, allowing the LLM to perform long-context reasoning with minimal computational overhead.

\begin{figure*}
    \centering
    \includegraphics[width=1.8\columnwidth]{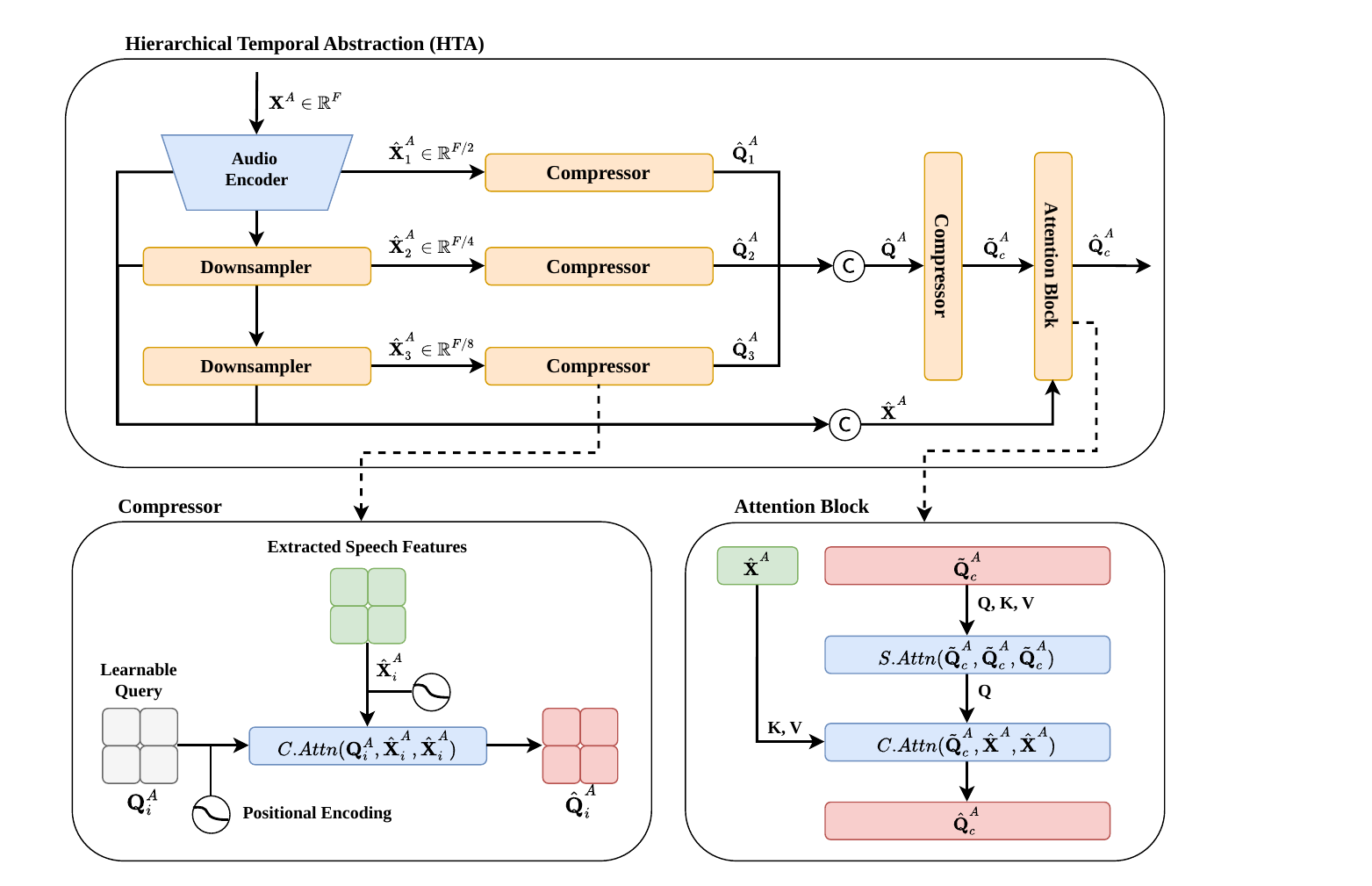}
    \caption{Detailed architecture of the Hierarchical Temporal Abstractor (HTA). $\text{C}$ denotes concatenation.}
    \label{fig:HTA}
\end{figure*}

\textbf{HTA (Hierarchical Temporal Abstractor)}: The HTA adopts a three-stage hierarchical design to progressively abstract speech information from acoustic to semantic levels. This progressive abstraction effectively distills continuous acoustic signals into compact semantic representations. The detailed structure of HTA is shown in Fig. \ref{fig:HTA}.

\textbf{Stage 1 (Local Acoustic Extraction)}: The first stage focuses on preserving fine-grained local information. The high-frame-rate features $\hat{\mathbf{X}}_1^A \in \mathbb{R}^{F}$ from the encoder are fused with a set of learnable queries $\mathbf{Q}_1^A$ via cross-attention. Here, $\mathbf{Q}_1^A$ acts as a local probe, extracting phoneme-level details while maintaining the original temporal resolution. To preserve order during this dense interaction, we inject sinusoidal positional encoding $PE(\cdot)$ \citep{positional} into both queries and keys. The compression operation at stage $i$ is defined as:

\begin{equation}
\begin{aligned}
\text{Compressor}(\mathbf{Q}_i^A, \hat{\mathbf{X}}_i^A, \hat{\mathbf{X}}_i^A)
&=
\text{Softmax}\!\left(
\frac{
\tilde{\mathbf{Q}}_i^A
\tilde{\mathbf{X}}_i^{A\!T}
}{\sqrt{d}}
\right)\\
\cdot \hat{\mathbf{X}}_i^A, \text{where } \tilde{\mathbf{Q}}_i^A &= \mathbf{Q}_i^A + PE(\mathbf{Q}_i^A), \\
\tilde{\mathbf{X}}_i^A &= \hat{\mathbf{X}}_i^A + PE(\hat{\mathbf{X}}_i^A), \\
\end{aligned}
\end{equation}

\textbf{Stages 2 \& 3 (Temporal Aggregation \& Semantic Abstraction)}: Stage 1 captures local acoustic details, while Stages 2 and 3 progressively abstract this information into higher-level semantic representations. A Downsampler acts as a temporal filter, reducing redundancy (e.g., silence, stationary noise). Each Downsampler consists of dual convolutional layers with kernel size 3 and GELU activation \citep{GELU}, applying a stride of 2 to halve temporal resolution at each step. Learnable queries $\mathbf{Q}_i^A$ at deeper stages attend to broader temporal contexts, transitioning from phoneme-level (Stage 1) to word/phrase-level (Stage 2), and finally to sentence-level semantics (Stage 3). The hierarchical feature update is formalized as:

\begin{equation}
\begin{split}
\hat{\mathbf{Q}}_i^A
= \text{Compressor}_i(
\mathbf{Q}_i^A,
\hat{\mathbf{X}}_i^A,
\hat{\mathbf{X}}_i^A), \\
\text{where} \quad
\hat{\mathbf{X}}_i^A
= \text{Downsampler}_i(
\hat{\mathbf{X}}_{i-1}^A).
\end{split}
\end{equation}

where $i \in \{2,3\}$, and $\hat{\mathbf{X}}_0^A$ denotes the initial encoder output. Multi-scale representations are concatenated as $\hat{\mathbf{Q}}^A = [\hat{\mathbf{Q}}_1^A; \hat{\mathbf{Q}}_2^A; \hat{\mathbf{Q}}_3^A]$ to form a comprehensive feature bank.

\textbf{Semantic Distillation (Extreme Token Compression)}: 
Although $\hat{\mathbf{Q}}^A$ contains rich information, its length scales linearly with audio duration, which remains suboptimal for long-form reasoning. To achieve highly compact speech representations, we introduce a final Semantic Distillation step.
Drawing inspiration from LLaVA-mini \citep{llava-mini}, which compresses visual features into a single token, we assume that speech segments can also be represented by a highly compact set of latent variables. However, unlike static images, speech represents a continuous, time-varying temporal stream. Therefore, instead of a single token, we employ a small, fixed number of learnable queries $\mathbf{Q}_c^A$ to distill the hierarchical information $\hat{\mathbf{Q}}^A$ into the final compressed tokens $\tilde{\mathbf{Q}}_c^A$:

\begin{equation}
\tilde{\mathbf{Q}}_c^A = \text{C.Attn}\left({\mathbf{Q}}_c^A, \hat{\mathbf{Q}}^A, \hat{\mathbf{Q}}^A\right).
\end{equation}

\textbf{Detail Recovery Mechanism}: 
Extreme compression risks losing subtle but critical acoustic cues (e.g., phonetic nuances, fine-grained acoustic transitions). To mitigate this, we incorporate a Detail Recovery attention block. This module allows the compressed semantic tokens $\tilde{\mathbf{Q}}_c^A$ to explicitly re-attend to the original multi-scale acoustic features $\hat{\mathbf{X}}^A$ and recover fine-grained acoustic information that may be attenuated during compression before entering the LLM.

\begin{equation}
\hat{\mathbf{Q}}_c^A = \text{Attention Block}\left(\tilde{\mathbf{Q}}_c^A, \hat{\mathbf{X}}^A\right), 
\end{equation}
where $\hat{\mathbf{X}}^A = [\hat{\mathbf{X}}_0^A; \hat{\mathbf{X}}_1^A; \hat{\mathbf{X}}_2^A]$.
The final output $\hat{\mathbf{Q}}_c^A$ achieves an effective balance between semantic density and acoustic fidelity, helping bridge the granularity mismatch described in Section \ref{sec:intro}. 

To substantiate the efficacy of our hierarchical compression mechanism, we provide a qualitative analysis of cross-attention patterns in Appendix~\ref{sec:appendix_analysis}. The visualizations show that HTA progressively shifts its attention toward deeper stages as speech duration increases, illustrating how the proposed hierarchical architecture facilitates effective long-range temporal abstraction. Furthermore, we present additional analyses on hierarchical temporal modeling in Appendix~\ref{sec:appendix_stagetoken} and investigate the impact of the intermediate query size ($\hat{\mathbf{Q}}_{i}^{A}$) in Appendix~\ref{appendix:q_size}.

\subsection{Three-Stage Training Strategy}
\label{sec:training}
To train FastSLM, we propose a three-stage speech modality adaptation strategy, designed to progressively enhance the model's capability to understand and adapt to speech input. Across all stages, we adopt low-rank adaptation (LoRA) \citep{LoRA} to ensure cost-efficient training with minimal trainable parameters. Specifically, we set the LoRA rank to 16 and alpha to 64, resulting in a scaling factor of 4.

\textbf{Pre-training (Short-form Speech Adaptation)}: In the first stage, the model is trained to adapt to short-form speech inputs. We construct a dataset of approximately 17K hours of speech-text pairs in both Korean and English, with each speech clip restricted to under 30 seconds. This ensures that the model can learn general ASR capabilities and effectively align speech with language. We adopt prompt formats inspired by hierarchical tags \citep{qwen-audio,qwen2-audio} to improve language-specific understanding during this stage. A detailed description can be found in Appendix \ref{sec:hie_tag}.

\textbf{Long-form Speech Adaptation}: Pre-trained speech encoders \citep{Conformer,Whisper,CLAP,Beats} are typically limited to processing segments shorter than 30 seconds, hindering performance on long-speech tasks such as speech summarization (SSUM) and spoken query-based question answering (SQQA). While Audio-Flamingo3 \citep{AF3} addressed this by constructing bespoke instruction tuning datasets, such an approach is time-consuming and costly.

To provide a cost-effective alternative, we train the model on a curated ASR-based dataset containing speech--text pairs ranging from 1 to 15 minutes. Rather than directly training abstract reasoning, this stage improves the model's ability to process extended speech sequences and maintain temporal coherence over long contexts. Since ASR datasets are more accessible than specialized instruction-tuning datasets, this approach provides a scalable path toward long-form speech adaptation.

\textbf{Instruction Tuning}: In the final stage, we perform instruction tuning to enable the model to handle a variety of downstream tasks. Due to the scarcity of non-English multi-task datasets, we generated a Korean multi-task dataset using a text-to-speech (TTS) engine \citep{melo_tts}, covering tasks such as SSUM and SQQA. Unlike previous stages, hierarchical language tags are no longer required as language identification is established; however, hierarchical task tags are employed to explicitly specify the task. Details are provided in Appendix \ref{sec:hie_tag}.

\section{Experiment Results}

\subsection{Dataset Description}
\textbf{Pre-training Dataset}: As described in Section 3.2, we constructed a bilingual dataset comprising 17K hours speech-text pairs to adapt ASR capabilities to the LLM during the pre-training stage. Including 9,152 hours of English speech-text pair (LibriSpeech \citep{Librispeech}, GigaSpeech-L \citep{Gigaspeech}, Voxpopuli \citep{Voxpopuli}, SpgiSpeech-M \citep{Spgispeech}, Earnings-22 \citep{Earnings-22}, AMI \citep{AMI}, Common Voice 15 \citep{common_voice}, AI-HUB ASR-En \citep{AIHUB}) dataset, and 7,812 hours of Korean speech-text pair (AI-Hub-ASR-Ko \citep{AIHUB}) dataset. A detailed description of the pre-training dataset can be found in Appendix \ref{appendix:pt} Table \ref{tab:pt-dataset}.

\textbf{Long-form speech Dataset}: 
To enhance the model capacity to process long-form speech input, particularly for tasks such as SSUM and SQQA, we constructed a dedicated long-form speech dataset. For English, we curated a total of 1,224 hours of long-form speech. For Korean, we curated a total of 1,012 hours of long-form speech.

\textbf{Instruction Tuning \& Evaluation Datasets}: To enable robust instruction-following capabilities, we constructed a multi-task instruction tuning dataset covering four representative speech-language tasks: ASR, automatic speech translation (AST), SSUM, and SQQA. For ASR, we randomly sampled Korean and English speech-text pairs from the during pre-training. For English SSUM, our dataset includes 1,600 hours of long-form dialogue from the MNSC corpus \citep{audio-llm}. 

Our English evaluations are conducted on natural human speech, providing evidence of FastSLM's robustness under realistic acoustic conditions. However, publicly available Korean speech datasets paired with high-quality semantic annotations for summarization and question answering remain extremely limited. Therefore, we synthesized Korean datasets by applying a TTS engine to the KMSS text summarization dataset \citep{Korean_SSUM} and KorQuAD dataset \citep{KoreanSQA}. We utilized TTS strictly as a controlled environment to evaluate cross-lingual semantic reasoning for Korean SSUM and SQQA, complementing our real-world English evaluations. A detailed breakdown of datasets used for each instruction tuning task is provided in Table~\ref{tab:it-dataset}.

\begin{table*}[!ht]
    \caption{Details of the instruction tuning dataset. ``En" denotes English, ``Ko" denotes Korean, and ``En2Ko", ``Ko2En" indicate the translation directions.}
    \label{tab:it-dataset}
    \centering
    \small
    \resizebox{1.4\columnwidth}{!}{
        \begin{tabular}{c | c c c c }
             \hline\hline
             Task & Dataset & Duration (hours) & \#Samples & speech Language \\ 
             \hline 
            \makecell[c]{ASR} & \makecell[c]{LibriSpeech \\ GigaSpeech-S \\ AI-HUB ASR} &
            \makecell[c]{960 \\ 250 \\ 1500} &
            \makecell[c]{281,241 \\ 230,068 \\ 320,000} &
            \makecell[c]{En \\ En \\ Ko }\\ [1ex]
            \hline
            \makecell[c]{AST} &
            \makecell[c]{AI-HUB AST (En2Ko) \\ AI-HUB AST (Ko2En)} &
            \makecell[c]{1,209 \\ 1,152} &
            \makecell[c]{400,000 \\ 400,000} &
            \makecell[c]{En \\ Ko} \\
            \hline
            \makecell[c]{SSUM} &
            \makecell[c]{SDS-PART6 \\ KMSS} &
            \makecell[c]{1,600 \\ 668} &
            \makecell[c]{103,935 \\ 84,000} &
            \makecell[c]{En \\ Ko} \\
            \hline
            \makecell[c]{SQQA} &
            \makecell[c]{LibriSQA \\ KorQuAD-speech} &
            \makecell[c]{364 \\ 483} &
            \makecell[c]{104,014 \\ 100,243} &
            \makecell[c]{En \\ Ko} \\  
            \hline
            Total & - & 8,186 & 2,023,501 & - \\
            \hline\hline
        \end{tabular}
    }
\end{table*}

To evaluate the speech understanding capabilities of FastSLM, we conducted experiments across a variety of benchmark tasks. 
\begin{itemize}
    \item \textbf{ASR}: For English, we used the OpenASR evaluation datasets \citep{openasr}. For Korean, we used the Common Voice 15 \citep{common_voice} and Fleurs \citep{Fleurs} datasets, which are open datasets, for fair comparison of results. We evaluate transcription quality using character error rate (CER) for Korean and word error rate (WER) for English to reflect the linguistic characteristics of each language. 
    \item \textbf{AST}: We evaluated translation on the Fleurs (En2Ko/Ko2En) \citep{Fleurs} and Minds14 (Ko2En) \citep{Minds14} datasets. We measured translation performance using the BLEU score \citep{BLEU}.   
    \item \textbf{SSUM}: Evaluation was conducted on SDS-PART6 \citep{sds6} and KMSS-speech \citep{Korean_SSUM}. Summarization quality was assessed using GPT-4 scoring with the LLM-as-a-judge framework \citep{LLM_judge}. To ensure evaluation fairness and prevent prompt engineering bias, we strictly adopted the zero-shot evaluation prompt and scoring rubric proposed in recent work \citep{phi4-m}. Please refer to Appendix \ref{appendix:prompt} for the exact SSUM judge prompt.
    \item \textbf{SQQA}: We measured accuracy on the LibriSQA and KorQuAD-speech datasets to evaluate SQQA performance.
\end{itemize} 

\subsection{Experimental Setup}
\textbf{Model Architecture}: FastSLM employs the encoder from Whisper-large-v3 \citep{Whisper} for speech feature extraction and adopts Qwen3-4B \citep{qwen3} as the backbone LLM for text generation. Despite its relatively compact size, Qwen3-4B exhibits sufficient capacity for comprehending speech-derived representations. In contrast to prior SLMs that typically utilize LLM backbone with 7 to 14 billion (B) parameters \citep{qwen-audio, qwen2-audio, salmonn, seg-former, AudioPaLM,Kimi-audio,Voxtral}, FastSLM achieves a favorable cost-performance trade-off by leveraging lightweight architecture without compromising performance \citep{phi4-m,AF2}. The HTA module within FastSLM compresses frame-level features via a hierarchical query-based mechanism. The number of learnable queries of $\mathbf{Q}_i^A$ was set to 80 cost-effectively \citep{seg-former}, and the number of learnable queries of $\mathbf{Q}_c^A$ used as contextual input in LLM was set to 50 through the experiment. For further ablation studies and justifications for parameter selection, please refer to Section \ref{ablation:token_num} Fig. \ref{fig:token_num}.

\textbf{Training}: FastSLM was trained on an NVIDIA A100 GPU-80GB$\times 4$ with a global batch size of 256. We used mixed precision training \citep{MPT} to maintain model performance while improving computational efficiency, with BF16 used as data type.

The model implementation details and the training setup are summarized in Appendix \ref{appendix:training}.

\subsection{Comparison with Baseline Speech Token Compression Methods}
To evaluate the effectiveness of HTA, we compare it with two representative query-based speech compression methods: the Segment-level Q-Former (SQ-Former)~\citep{seg-former}, which compresses speech at the segment level, and the Window-level Q-Former (WQ-Former)~\citep{salmonn}, which performs compression within local temporal windows. We further include a simple average pooling (AvgPool) baseline, where frame-level speech features are temporally downsampled via AvgPool and projected into the LLM embedding space using a multi-layer perceptron (MLP). This baseline serves to evaluate whether straightforward temporal downsampling can provide an effective alternative to learnable query-based compression.

We evaluate all methods using the ASR task, as it provides a direct measure of how accurately the model can understand speech content. 
In addition, to assess the computational load imposed on the LLM when processing long-form speech, we measure the LLM computational cost in terms of FLOPs using a 5-minute speech input. 
All baselines were trained and evaluated under the same pre-training dataset, LLM backbone, LoRA configuration, and embedding dimensions to ensure a fully fair comparison. 
The detailed results are presented in Table~\ref{tab:baseline}.

\begin{table}[!h]
    \caption{\small{Comparison of ASR performance and computational complexity across baseline compression methods. LS denotes the LibriSpeech.}}
    \label{tab:baseline}
    \centering
    {\footnotesize
    \resizebox{\columnwidth}{!}{
        \begin{tabular}{c | c c c c}
             \hline\hline
             \diagbox[width=11em]{Dataset}{Method} & AvgPool & SQ-Former & WQ-Former & HTA (ours) \\
             \hline
             LS-clean & $1.88$ & $2.32$ & $2.14$ & $2.09$ \\ [0.5ex]
             LS-other & $4.12$ & $4.87$ & $4.51$ & $4.67$ \\ [0.5ex]
             Voxpopuli & $7.14$ & $8.37$ & $7.26$ & $6.55$ \\ [0.5ex]
             \hline
             \makecell[c]{\#Speech Tokens/Sec.} & $25.0$ & $2.67$ & $2.93$ & $\mathbf{1.67}$ \\ [0.5ex]
             \makecell[c]{LLM FLOPs (T)} & $30.6$ & $3.32$ & $3.65$ & $\mathbf{2.15 (-92\%)}$ \\ [0.5ex]
            \hline\hline
        \end{tabular}
    }}
\end{table}

As shown in Table~\ref{tab:baseline}, HTA achieves the best WER on VoxPopuli while maintaining competitive performance on LibriSpeech with the lowest token rate (1.67 tokens/sec). In contrast, AvgPool requires substantially more speech tokens (25 tokens/sec) and 14.2$\times$ higher LLM computation (30.6 vs. 2.15 TFLOPs) for only marginal gains on LibriSpeech, demonstrating the effectiveness of HTA for efficient long-form speech processing.

\begin{table*}[!h]
    \caption{Comparison of FastSLM with other SLMs on various tasks. Underline indicates the second-best performance. WER and CER are lower-is-better ($\downarrow$); Accuracy (ACC), BLEU, and Score are higher-is-better ($\uparrow$). N/A indicates the model lacks the corresponding capability. `*' denotes results fine-tuned on an additional Korean dataset provided in the official Hugging Face supplementary material. Detailed ASR benchmark WER for each dataset is reported in Appendix \ref{sec:asr}.}
    \label{tab:sota}
    \centering
    \small
    {
    \resizebox{1.8\columnwidth}{!}{
        \begin{tabular}{c | c c c c c c c c c}
             \hline\hline
             Task & Metric & Dataset & \makecell[c]{FastSLM \\ 4.8B} & \makecell[c]{Whisper \\ 1.5B}  & \makecell[c]{Qwen2-Audio \\ 8B}  & \makecell[c]{Phi4-MM \\ 5.8B} & \makecell[c]{ Voxtral-mini \\ 4.7B} & \makecell[c]{Gemini-2.5- \\ Flash}\\
             \hline
             \multicolumn{3}{c|}{\text{\#speech tokens/30 Sec.}} & 
             50 & 1500 & 101 & 375 & 750 & 960 \\
            \hline
             \makecell[c]{ASR \\ (En)} & WER $\downarrow$ & OpenASR & $\underline{6.47}$ & $7.44$ & $7.43$ & $\mathbf{6.14}$  & $7.05$ & $9.29$ \\ [0.5ex]
             \makecell[c]{ASR \\ (Ko)} & CER $\downarrow$ & \makecell[c]{Fleurs \\ Common Voice 15} & $\mathbf{3.82}$ & $7.92$ & N/A & N/A & N/A & $\underline{4.55}$ \\ [0.5ex]
             \hline
             \makecell[c]{AST \\ (En2Ko)} & BLEU $\uparrow$ & \makecell[c]{Fleurs} &
             $\underline{7.39}$ & N/A & N/A & *$2.62$ & N/A & $\mathbf{13.4}$\\ [0.5ex]
             \makecell[c]{AST \\ (Ko2En)} & BLEU $\uparrow$ & \makecell[c]{Fleurs} &
             $\mathbf{19.5}$ & $18.6$ & N/A & *$10.4$ & N/A & $\underline{19.2}$ \\ [0.5ex]
             \makecell[c]{AST \\ (Ko2En)} & BLEU $\uparrow$ & \makecell[c]{Minds14} &
             $\underline{28.9}$ & $\mathbf{29.5}$ & N/A & *$14.8$ & N/A & $26.3$\\ [0.5ex]
             \hline
             \makecell[c]{SSUM \\ (En)} & Score (1-7) $\uparrow$ & \makecell[c]{SDS-PART6} &
             $5.40$ & N/A & $4.54$ & $5.30$ & $\underline{5.48}$ & $\mathbf{5.87}$\\ [0.5ex]
             \makecell[c]{SSUM \\ (Ko)} & Score (1-7) $\uparrow$ & \makecell[c]{KMSS} &
             $\underline{4.12}$ & N/A & N/A & N/A & N/A & $\mathbf{4.37}$ \\ [0.5ex]
             \hline
             \makecell[c]{SQQA \\ (En)} & ACC $\uparrow$ & \makecell[c]{LibriSQA} & 
             $\mathbf{69.5}$ & N/A & $57.2$ & $64.5$ & $48.9$ & $\underline{67.0}$ \\ [0.5ex]
             \makecell[c]{SQQA \\ (Ko)} & ACC $\uparrow$ & \makecell[c]{KorQuAD- \\ speech} & $\mathbf{64.9}$ & N/A & N/A & N/A & N/A & $\underline{64.8}$ \\ [0.5ex]
            \hline\hline
        \end{tabular}
    }}
\end{table*}

In contrast, HTA achieves a strong balance between accuracy and efficiency: it reduces the token rate by 38\% compared to WQ-Former and lowers LLM FLOPs by 41.1\% (3.65T → 2.15T), while still improving WER on VoxPopuli and LS-clean. These results indicate that HTA provides a significantly more efficient speech-to-LLM alignment mechanism, greatly reducing the computational burden of autoregressive decoding for long-form speech while maintaining competitive recognition performance.

\subsection{Quantitative Results}

We primarily compare FastSLM with strong speech-centric baselines such as WhisperV3 and AST, which directly align with our speech-only setting. For completeness, we additionally evaluate several multimodal models (Qwen2-Audio \citep{qwen2-audio}, Phi-4-Multimodal (MM) \citep{phi4-m}, Gemini-2.5-Flash \citep{gemini}, and Voxtral-Mini \citep{Voxtral}) in their speech-only mode. 

We note that comparisons between specialized speech models and massive general multimodal LLMs (e.g., Gemini-2.5-Flash, which is trained on extensive multilingual datasets heavily including Korean) should be interpreted cautiously due to differences in training data scale and optimization objectives. These broader comparisons are intended to demonstrate the practical competitiveness of FastSLM rather than strict absolute architectural superiority. 

FastSLM demonstrates a powerful combination of efficiency and performance, achieving competitive results with just 50 speech tokens per 30-second input—a fraction of that used by models like Whisper (1,500) and Gemini-2.5-Flash (960). As detailed in Table~\ref{tab:sota}, its key achievements include:
\begin{itemize}
    \item \textbf{ASR}: Achieves a highly competitive CER of 3.82 on Korean benchmarks and a strong WER of 6.47 on English OpenASR, outperforming several larger models.
    \item \textbf{AST}: Demonstrates the highest performance among evaluated models on the Fleurs Ko2En task with 19.5 BLEU, outperforming both Whisper and Gemini-2.5-Flash. Furthermore, it shows competitive performance on the Minds14 dataset (28.9 BLEU), comparable to the strong Whisper baseline.
    \item \textbf{SSUM}: Delivers competitive scores of 5.40 (English) and 4.12 (Korean), performing on par with several larger SLMs despite the extreme token compression.
    \item \textbf{SQQA}: Achieves robust accuracy on the evaluated benchmarks, attaining 69.5\% on LibriSQA (English) and 64.9\% on KorQuAD-speech (Korean), establishing its strong reasoning capabilities.
\end{itemize}

In summary, FastSLM demonstrates that hierarchical speech abstraction enables efficient long-form speech understanding while maintaining competitive performance across diverse speech-language tasks.

\subsection{Ablation Study}

\label{ablation:token_num}
\begin{figure}[!h]
    \centering
    \includegraphics[width=0.9\columnwidth]{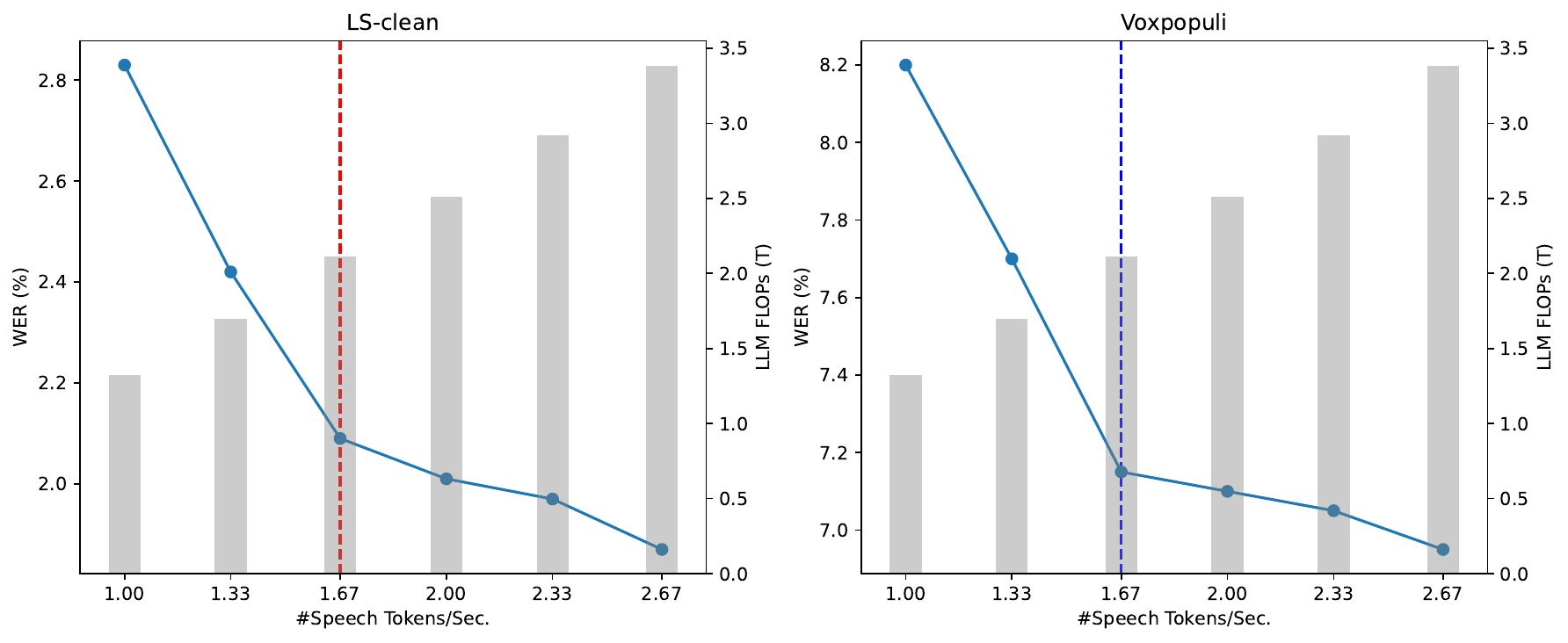}
    \caption{\small{ASR performance of FastSLM with various speech tokens. (left) LS-clean decoding result, and (right) Voxpopuli decoding result.}}
    \label{fig:token_num}
\end{figure}

\textbf{Effect of Speech Token Compression Ratio on ASR Performance}:
To determine the optimal speech token compression ratio, we conducted an ablation study that evaluates the trade-off between ASR performance (WER) and computational cost. As illustrated in Fig. \ref{fig:token_num}, a clear relationship emerges. While a high token rate (2.67 tokens/sec) yields the best ASR performance, it incurs a substantial computational cost. In contrast, an overly compressed representation (1.33 tokens/sec) results in a significant degradation of performance. 

To rigorously justify our effective operating point rate of 1.67 tokens/sec, we analyze the capacity of the HTA module through the lens of information bottleneck theory. We quantify the trade-off between reasoning performance and computational cost by defining the \textit{Marginal Semantic Utility} ($\eta$). This metric measures the effective semantic gain per unit of computational overhead:

\begin{equation}
\eta(r) = \left| \frac{\partial \text{WER}(r)}{\partial \text{FLOPs}(r)} \right| \approx \frac{\Delta \text{WER}}{\Delta \text{FLOPs}}
\end{equation}

where $r$ represents the speech token rate. We apply this metric to the LS-clean decoding results, as summarized in Table~\ref{tab:efficiency_score}.

\begin{table}[!h]
    \caption{\small{Marginal Semantic Utility ($\eta$) across different temporal compression intervals. A sharp drop indicates semantic information saturation.}}
    \label{tab:efficiency_score}
    \centering
    \small
    \resizebox{\columnwidth}{!}{
        \begin{tabular}{c | c c c}
             \hline\hline
             Token Rate Interval ($r$) & $\Delta$WER & $\Delta$FLOPS & Utility ($\eta$) $\uparrow$\\
             \hline
             $1.33 \rightarrow 1.67$ & $\approx 0.3\%$ & $\approx 0.3$ T & $\approx 1.00$ \\
             $1.67 \rightarrow 2.00$ & $\approx 0.1\%$ & $\approx 1.3$ T & $\approx 0.08$ \\
            \hline\hline
        \end{tabular}
    }
\end{table}

The analysis reveals a precipitous drop in utility—from 1.00 to 0.08—when expanding the token rate beyond 1.67 tokens/sec. This non-linear degradation demonstrates that the interval beyond 1.67 tokens/sec marks a stark point of diminishing returns. In the $1.33 \rightarrow 1.67$ interval, additional tokens actively resolve linguistic ambiguities, yielding high utility. Conversely, beyond this bottleneck, additional tokens merely encode redundant acoustic variations (e.g., micro-pauses), quadratically inflating the LLM's computational cost without improving reasoning. Crucially, this empirical saturation point aligns with typical human speaking rates of approximately 100--120 words per minute (approximately 1.67 to 2.0 words/sec) observed in spontaneous speech \citep{speech_rate}, effectively synchronizing the acoustic stream with the discrete semantic processing pace of the LLM.

\begin{figure*}[!ht]
    \centering
    \includegraphics[width=2.0\columnwidth]{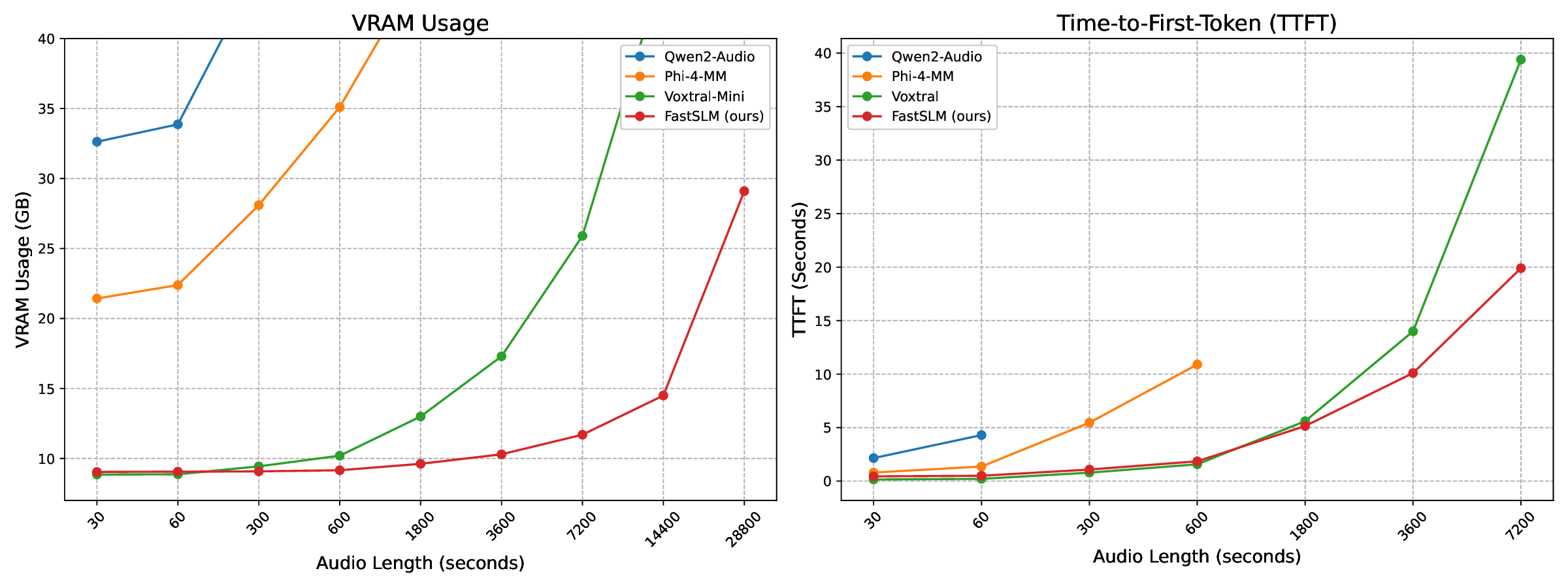}
    \caption{\small{Comparison VRAM usage and time-to-first-token (TTFT) according to speech length.}}
    \label{fig:scaling}
\end{figure*}

\textbf{Scaling Limits of Long-Form Speech Input:}
To provide a practical assessment beyond indirect metrics like FLOPs \citep{shufflenetv2}, we empirically evaluated FastSLM's scaling properties by measuring VRAM consumption and time-to-first-token (TTFT) on a single 40GB A100 GPU.

As shown in Fig. \ref{fig:scaling}, FastSLM demonstrates significant scalability advantages. While benchmark models exhibit exponential VRAM growth—and models like Qwen2-Audio or Phi-4-MM suffer from Out-of-Memory (OOM) failures when processing audio beyond several minutes—FastSLM scales near-linearly, successfully processing an 8-hour speech stream using under 30GB of memory. Furthermore, FastSLM maintains a competitive TTFT that increases only minimally with speech length, avoiding the sharp latency spikes observed in similarly-sized baselines. We acknowledge that recognition accuracy eventually degrades on extremely long out-of-distribution audio due to our training data limits, but this empirical result explicitly proves that the structural scalability of the architecture (OOM prevention) remains highly robust.

\textbf{Impact of Hierarchical Modules and Training Strategy:} To validate the efficacy of our architectural design and the three-stage training strategy, we conducted an ablation study by selectively removing key components: the Downsampler stage, the hierarchical attention mechanism, and Training Stage 2 (long-form speech adaptation). 

\begin{table}[!h]
    \caption{\small{Comparison of long-form speech adaptation strategy.}}
    \label{tab:training_stage}
    \centering
    \small
    {
    \resizebox{\columnwidth}{!}{
        \begin{tabular}{c | c c c}
             \hline\hline
             \diagbox[width=12em]{Method}{Dataset} & \makecell[c]{LS-Long \\ WER $\downarrow$} & \makecell[c]{KorQuAD-Speech \\ ACC $\uparrow$}  & \makecell[c]{SDS-PART6 \\ Score (1-7) $\uparrow$}\\
             \hline
             \makecell[c]{w/o Downsample Stage} & $12.4$ & $56.7$ & $4.12$ \\ [0.5ex]
             \makecell[c]{w/o Hierarchical Attention} & $10.8$ & $56.9$ & $4.92$ \\ [0.5ex]
             \makecell[c]{w/o Training Stage 2} & $6.81$ & $59.0$ & $5.07$ \\ [0.5ex]
             FastSLM & $\mathbf{5.78}$ & $\mathbf{64.9}$ & $\mathbf{5.40}$ \\ [0.5ex]
            \hline\hline
        \end{tabular}
    }}
\end{table}

As summarized in Table~\ref{tab:training_stage}, removing the Downsampler or hierarchical attention caused a severe performance degradation, with WER on LS-Long \citep{libri_long} surging to 12.4 and 10.8, respectively. This underscores the structural necessity of our hierarchical compression in handling long sequences. Furthermore, omitting Training Stage~2, which aligns the model with long-context speech via Long-form speech adaptation, resulted in significant regression across tasks—specifically, increasing LS-Long WER from 5.78 to 6.81 and dropping KorQuAD-Speech accuracy from 64.9\% to 59.0\%. The full FastSLM configuration consistently outperforms all ablated variants, confirming that both the hierarchical architecture and the dedicated adaptation stage are integral to achieving efficient and robust long-form understanding.

\section{Conclusion}
In this paper, we introduced FastSLM, a lightweight architecture designed to overcome the scaling bottlenecks of long-form speech processing. By employing the HTA and a cost-effective three-stage training strategy, FastSLM compresses high-frame-rate acoustic features into compact semantic tokens—reducing sequence length by up to 97\%. Our results demonstrate that this extreme token compression achieves competitive performance across diverse speech-language benchmarks, providing a highly scalable and efficient framework for long-form speech-language modeling.

\section*{Limitations}
FastSLM has several limitations. First, due to the scarcity of publicly available Korean long-form datasets, Korean instruction tuning relies on TTS-generated speech, which may introduce a synthetic-to-real domain gap. Second, our evaluation focuses on linguistic speech understanding (ASR, AST, SSUM, and SQQA); preserving paralinguistic information such as emotion, speaker identity, or environmental sounds remains future work. Third, while FastSLM can efficiently process multi-hour audio without memory bottlenecks, recognition accuracy on extremely long recordings may further benefit from larger-scale long-form training data.

\section*{Ethical Considerations}
FastSLM is trained exclusively on publicly available ASR corpora and synthetic data, avoiding the use of proprietary or sensitive personal audio. As with other speech-language models, it should be deployed responsibly to mitigate potential misuse, such as unauthorized large-scale speech surveillance. We also acknowledge that demographic biases in the training data may affect performance across accents and speaking styles. The proposed token-efficient architecture substantially reduces the computational cost of long-form speech processing, contributing to more sustainable AI deployment.

\section*{Acknowledgments}
This work was supported by the Institute of Information \& Communications Technology Planning \& Evaluation(IITP) grant funded by the Korea government(MSIT)(No.RS-2026-25508985, Technology Development for a Digital Platform for Integrated Disaster Management in Super-High-Rise Complex Facilities).

\bibliography{custom}

\appendix
\onecolumn
\section{Qualitative Analysis of HTA Attention Map}
\label{sec:appendix_analysis}

\begin{figure*}[h!] 
    \centering     
    \caption{
        Visualization of HTA attention patterns across speech duration (logarithmic scale). 
        Top: short-form speech ($<30\,\mathrm{s}$). 
        Middle: mid-form speech ($<60\,\mathrm{s}$). 
        Bottom: long-form speech ($>15\,\mathrm{min}$).
    }
    \label{fig:attn_map}
    \includegraphics[width=0.8\columnwidth]{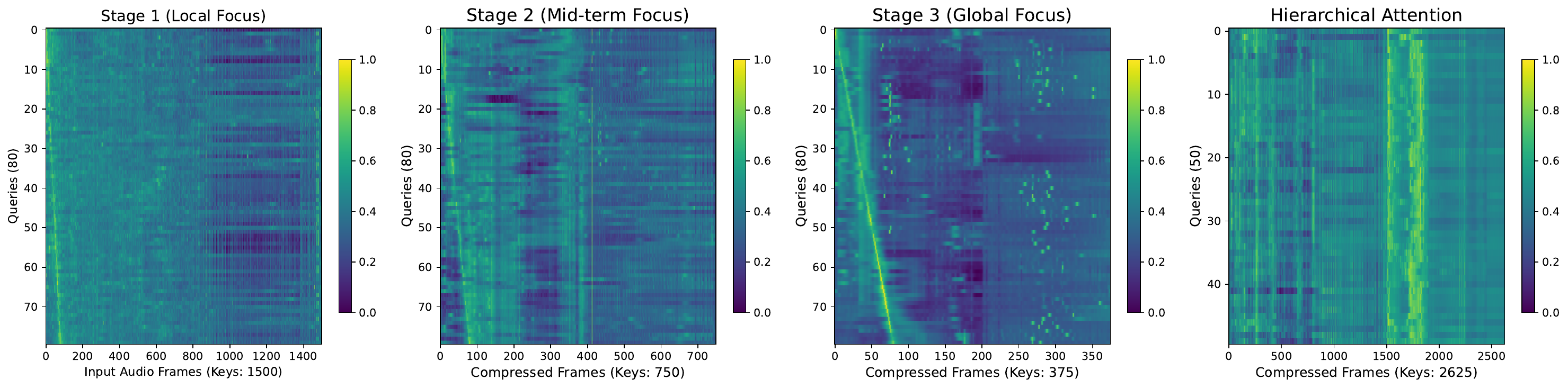}
    \\
    \vspace{3pt}
    \includegraphics[width=0.8\columnwidth]{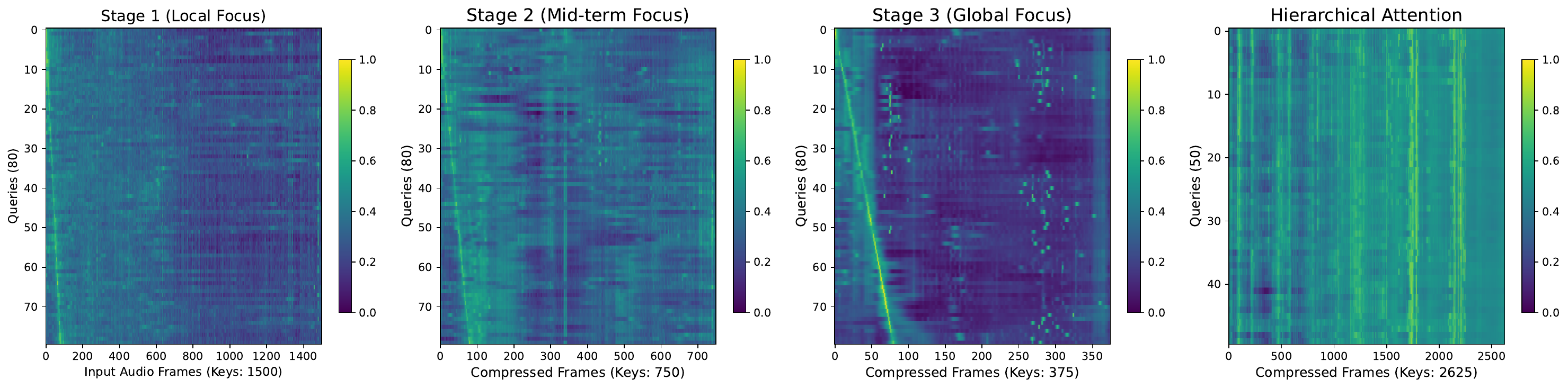}
    \\
    \vspace{3pt} 
    \includegraphics[width=0.8\columnwidth]{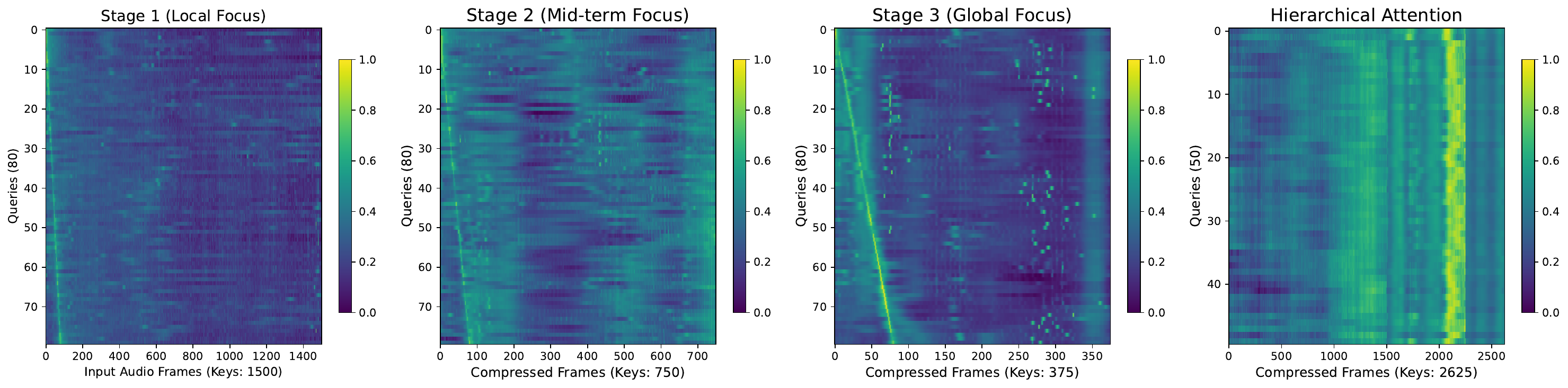}
\end{figure*}

Fig. \ref{fig:attn_map} shows the cross-attention patterns of HTA for short-form (top), mid-form (middle), and long-form (bottom) speech across the three hierarchical stages.

For short-form speech, the final queries attend broadly to all stages, indicating that local and mid-term features remain useful when the sequence is short.
For mid-form speech, attention begins to shift away from Stage~1 and is increasingly concentrated on Stage~2 and Stage~3, reflecting the need for broader temporal context.
For long-form speech, attention becomes strongly dominated by Stage~3, while Stage~1 and Stage~2 receive minimal attention. This shows that the model relies on high-level, compressed representations for long-form speech reasoning.

Overall, as speech length increases, the attention distribution progressively moves from local (Stage~1) to global (Stage~3) features, demonstrating that HTA adaptively adjusts its focus based on speech length.

\section{Additional Analysis of Hierarchical Temporal Modeling}
\label{sec:appendix_stagetoken}

To further validate the effectiveness of hierarchical temporal abstraction, we compare HTA with a larger single-stage abstractor under the same token budget. While the single-stage abstractor performs feature compression only once, HTA progressively aggregates speech representations across multiple temporal scales, enabling more effective modeling of both local acoustic details and long-range semantic context. Additional cross-attention visualizations (Appendix~\ref{sec:appendix_analysis}) further illustrate that HTA gradually shifts its attention toward deeper stages as the input duration increases. The quantitative comparison is presented in Table~\ref{tab:large_token}.

\begin{table}[h!]
    \centering
    \caption{Comparison between a larger single-stage abstractor and our HTA under the same token budget.}
    \vspace{3pt}
    \label{tab:large_token}
    \small
    \resizebox{0.5\columnwidth}{!}{
        \begin{tabular}{c|c c}
            \hline\hline
            \diagbox[width=10em]{Dataset}{Method} & \makecell[c]{Larger Single-stage \\ Abstractor} & HTA (ours) \\
            \hline
            \makecell[c]{LS-clean (WER $\downarrow$)} & 2.11 & \textbf{2.09} \\ [1ex]
            \makecell[c]{Voxpopuli (WER $\downarrow$)} & 7.11 & \textbf{6.55} \\ [1ex]
            \makecell[c]{SDS-PART6-Speech \\ Score (1-7) $\uparrow$}  & 5.04 & \textbf{5.40} \\ [1ex]
            \makecell[c]{KorQuAD-Speech \\ ACC $\uparrow$} & 61.2 & \textbf{64.9} \\ [0.5ex]
            \hline\hline
        \end{tabular}
    }
\end{table}

Although both models achieve comparable ASR performance on LS-clean, HTA consistently outperforms the larger single-stage abstractor on long-form and reasoning-intensive tasks, including VoxPopuli, SDS-PART6-Speech, and KorQuAD-Speech. These results indicate that progressive hierarchical abstraction is more effective than increasing the capacity of a single-stage compression module under the same token budget.

\section{Effect of Hierarchical Staging in HTA} 
\label{appendix:q_size}

To directly evaluate the benefit of the hierarchical design in HTA, we compare three structural variants: utilizing only Stage 1 (w/o Stage 2/3), utilizing Stages 1 and 2 (w/o Stage 3), and the full hierarchical pipeline (HTA). 

\begin{table}[h!]
    \centering
    \caption{Effect of hierarchical downsampling stages on speech understanding performance.}
    \vspace{3pt}
    \label{tab:stage_ablation}
    \small
    \resizebox{0.6\columnwidth}{!}{
        \begin{tabular}{c|c c c}
            \hline\hline
            \diagbox[width=8em]{Method}{Dataset} & \makecell[c]{LS-clean \\ WER $\downarrow$} & \makecell[c]{SDS-PART6-Speech \\ Score (1-7) $\uparrow$}  & \makecell[c]{KorQuAD-Speech \\ ACC $\uparrow$}\\            
            \hline
            w/o Stage 2/3 & 2.23 & 4.12 & 56.7 \\
            w/o Stage 3 & 2.15 & 4.98 & 62.2 \\
            HTA & \textbf{2.09} & \textbf{5.40} & \textbf{64.9} \\
            \hline\hline
        \end{tabular}
    }
\end{table}

As shown in Table~\ref{tab:stage_ablation}, performance consistently improves across all metrics as more downsampling stages are incorporated. While the foundational speech recognition capability (LS-clean WER) shows a steady reduction from 2.23 to 2.09, the most substantial enhancements are observed in long-form and reasoning-intensive tasks. Specifically, advancing from a single-stage to the full three-stage hierarchy significantly boosts the SSUM score (SDS-PART6-Speech) from 4.12 to 5.40 and increases the SQQA accuracy (KorQuAD-Speech) by an absolute 8.2\% (from 56.7\% to 64.9\%). These empirical results explicitly demonstrate that progressive temporal abstraction is essential for effectively capturing the global semantic context required in long-form speech processing.

\section{Prompt Template and Hierarchical Tags for Training}
\label{sec:hie_tag}
We applied a unified prompt template and task/language control tokens during training to support multiple speech-language tasks:

\begin{equation}
    \begin{gathered}
            \text{User}: <|\text{audio\_bos}|><|\text{AUDIO}|><|\text{audio\_eos}|> \\ \{\text{Prompt/Question}\} \text{\textbackslash n} \,\text{Assistant}:
    \end{gathered}
\nonumber
\end{equation}

To improve task specialization and language awareness, we employed hierarchical task and language tokens, which enabled robust detection and performance across languages and tasks:

\begin{equation*}
    \begin{gathered}
    \text{Language Token}: <|\text{KO}|><|\text{EN}|> \\
    \text{Task Token}: <|\text{ASR}|><|\text{AST}|> \\ <|\text{SQQA}|><|\text{SSUM}|>
    \end{gathered}
\end{equation*}

\section{Pre-training Dataset}
Table \ref{tab:pt-dataset} is a dataset used for pre-training of FastSLM. The dataset consists of English and Korean, and consists of a total of 10M speech-text pairs. 
\label{appendix:pt}

\begin{table*}[!h]
    \caption{Pre-training dataset details. En denote the English, and Ko denote the Korean.}
    \label{tab:pt-dataset}
    \centering
    \small
    \resizebox{0.7\columnwidth}{!}{
        \begin{tabular}{c c c c}
             \hline\hline
             Dataset & Duration (hours) & \# Samples & Speech Language \\ 
            \hline 
            LibriSpeech & 960 & 281,241 & En \\
            TED-LIUM-release3 & 454 & 268,263 & En \\
            GigaSpeech-L & 2,500 & 2,266,371 & En \\
            Voxpopuli & 523 & 182,482 & En \\ 
            SpgiSpeech-M & 1,000 & 385,361 & En \\
            Earnings-22 & 105 & 52,006 & En \\
            AMI & 78 & 108,502 & En \\
            Common Voice 15 & 2,532 & 1,070,066 & En \\
            AI-HUB ASR-En & 1,000 & 1,020,265 & En \\
            AI-HUB ASR-Ko & 7,812 & 4,557,512 & Ko \\
            \hline
            Total & 16,964 & 10,212,348 & - \\
            \hline\hline
        \end{tabular}
    }
\end{table*}

\section{Prompt for GPT-4 as a Judge on Speech Benchmarks}\label{sec:llm-as-a-judge}
\label{appendix:prompt}
\small
The following is the exact prompt template used for evaluating SSUM output via an LLM-as-a-Judge.

\begin{lstlisting}[style=prompt,caption={LLM-as-a-Judge Prompt for SSUM Evaluation}]
You are a skilled evaluator for summaries generated based on user-provided instructions.
Your task is to rate how well the summary follows the user's instructions on a 1-7 scale.

Scoring Rubric:
- 7 (Excellent): Fully follows all instructions. Accurate, fluent, and coherent with the correct level of detail and structure.
- 6 (Good): Almost perfect, with very minor issues that do not affect usability (e.g., tiny structural deviation, trivial omission).
- 5 (Mostly Correct): Fulfills the main instruction but has noticeable issues (e.g., includes some unimportant extras, misses a few details).
- 4 (Acceptable): Adheres to the instruction partially but has significant issues like inconsistencies or irrelevant content.
- 3 (Poor): Minimally adheres to the instruction, missing most required details or containing significant irrelevant/hallucinated content.
- 2 (Very Poor): Fails to follow the core instruction. Mostly irrelevant, fabricated, or ignores requested structure/tone.
- 1 (Fails): Completely fails to follow instructions.

Input:
- User Instruction: {USER_INSTRUCTION}
- Reference (gold): {REFERENCE_ANSWER}
- Model Summarization: {SUMMARY_TO_EVALUATE}

Notes:
- It helps to read the Summary first, then compare with the Reference and Instruction.
- If the summary is missing or empty, return N/A as the score.

Output:
Note: Use the following JSON format for easy downstream consumption.
{
    explanation: Brief reasoning for the score based on the rubric.,
    score: <Float, 1-7>
}
\end{lstlisting}

\vspace{-10pt}
\section{Generation Configuration}
We use the following decoding hyper-parameters for all LLM-based generation tasks.

\begin{table}[h!]
    \centering
    \small
    \caption{Decoding configuration used for LLM-based generation in FastSLM.}
    \label{tab:generation_config}
    \begin{tabular}{c|c}
    \hline\hline
        Parameter  &  Value \\
        \hline
        Decoding Strategy & Sampling \\
        Temperature & 0.2 \\
        Top-p & 0.95 \\
        Top-k & 20 \\
        Repetition Penalty & 1.0 \\
        \hline\hline
    \end{tabular}
\end{table}

\section{Model and Training Parameters}
\label{appendix:training}
The model implementation details and the training setup and  for each stage are presented in Table~\ref{tab:model_config} and Table~\ref{tab:training}.

\begin{table}[ht!]
    \centering
    \small
    \caption{Model configuration for FastSLM.}
    \label{tab:model_config}
    \resizebox{0.6\columnwidth}{!}{
    \begin{tabular}{c c c}
        \hline\hline
        Module & Component & Configuration \\
        \hline
        \makecell[c]{Encoder} & 
        \makecell[c]{Backbone \\ Parameters \\ Hidden Size \\ Context Length} &
        \makecell[c]{Whisper-large-v3 \\ 635M \\ 1280 \\ 1500} \\
        \hline

        \makecell[c]{Abstractor} &
        \makecell[c]{Backbone \\ Parameters \\ Hidden Size \\ Queries per Stage \\ Compressed Speech Token \\ Downsampling Factors} &
        \makecell[c]{HTA \\ 56M \\ 1280 \\ 80 \\ 50 \\ 2} \\
        \hline

        \makecell[c]{LLM} &
        \makecell[c]{Backbone \\ Parameters \\ Hidden Size \\ Context Length} &
        \makecell[c]{Qwen3-4B \\ 4.06B \\ 2560 \\ 4096} \\
        \hline

        \makecell[c]{LoRA} &
        \makecell[c]{Rank ($r$) \\ Alpha ($\alpha$) \\ Scaling Factor \\ LoRA Target Modules} &
        \makecell[c]{16 \\ 64 \\ 4 \\ q/k/v\_proj, gate/up/down\_proj} \\
        \hline\hline
    \end{tabular}
    }
\end{table}

\begin{table}[!h]
    \caption{Training settings across stages}
    \label{tab:training}
    \centering
    \small
    \resizebox{0.5\columnwidth}{!}{
        \begin{tabular}{c c c c}
             \hline\hline
             Setting & Stage1 & Stage2 & Stage3 \\ 
            \hline 
            Learning Rate & 1e-4 & 5e-5 & 5e-5 \\
            Learning Rate Scheduler & \multicolumn{3}{c}{Linear Decay} \\
            Weight Decay & 0 & 1e-4 & 1e-4 \\
            Epoch & 1 & 1 & 2 \\
            Data Type & \multicolumn{3}{c}{BF16} \\
            DeepSpeed Stage & \multicolumn{3}{c}{Zero2} \\
            \hline\hline
        \end{tabular}
    }
\end{table}

\section{Details of ASR Benchmark Results}
Table \ref{tab:OpenASR} presents a detailed comparison of FastSLM and SOTA models across multiple ASR benchmarks. 
We report WER for English datasets and CER for Korean datasets. 
The results demonstrate that FastSLM achieves competitive performance while using significantly fewer speech tokens per second.

\label{sec:asr}
\begin{table*}[!ht]
    \caption{Comparison of WER between FastSLM and state-of-the-art (SOTA) models. These results represent ASR Benchmark Dataset WER and CER.}
    \label{tab:OpenASR}
    \centering
    {\small
    \resizebox{\columnwidth}{!}{
        \begin{tabular}{c | c c c c c c c c}
             \hline\hline
             Dataset & Sub-Category & Metric & \makecell[c]{FastSLM \\ 4.8B} & \makecell[c]{Qwen2-Audio \\ 8B}  & \makecell[c]{Phi4-Multimodal \\ 5.8B} & \makecell[c]{Whisper \\ 1.5B} & \makecell[c]{ Voxtral-mini \\ 4.7B} & \makecell[c]{Gemini-2.5- \\ Flash}\\
             \hline
             \multirow{8}{*}{OpenASR} 
             & AMI & WER  & $10.8$ & $15.2$ & $11.7$ & $16.0$ & $16.3$ & $21.6$ \\ [0.5ex]
             & Earnings22 & WER  & $10.7$ & $14.1$ & $10.2$ & $11.3$ & $10.7$ & $13.1$ \\ [0.5ex]
             & GigaSpeech & WER  & $10.7$ & $10.3$ & $9.78$  & $10.0$ & $10.2$ & $10.7$ \\ [0.5ex]
             & SpgiSpeech & WER  & $2.33$ & $3.00$  & $3.13$  & $2.01$  & $2.37$  & $3.82$  \\ [0.5ex]
             & TEDLIUM  & WER    & $3.97$ & $4.05$  & $2.90$  & $3.91$  & $3.68$  & $3.01$  \\ [0.5ex]
             & LS-clean  & WER   & $2.09$ & $1.74$  & $1.68$  & $2.94$  & $1.88$  & $2.49$  \\ [0.5ex]
             & LS-other & WER    & $4.67$ & $4.03$  & $3.83$  & $3.86$  & $4.10$  & $5.84$  \\ [0.5ex]
             & Voxpopuli & WER   & $6.55$ & $7.05$  & $5.91$  & $9.54$  & $7.14$  & $7.89$  \\ [0.5ex]
             \hline
             \multirow{2}{*}{Fleurs}
             & En & WER  & $5.26$ & $5.27$ & $3.38$ & $4.10$ & $3.77$ & $6.20$\\ [0.5ex]
             & Ko & CER  & $2.79$ & N/A & N/A & $5.32$ & N/A & $3.00$\\ [0.5ex]
             \hline
             \multirow{2}{*}{Common Voice 15}
             & En & WER & $10.9$ & $8.68$ & $7.61$ & $9.30$ & $10.2$ & $11.2$ \\ [0.5ex]
             & Ko & CER & $4.55$ & N/A & N/A & $5.74$ & N/A & $6.09$\\ [0.5ex]
             \hline\hline
        \end{tabular}
    }}
\end{table*}

\section{Use of AI Writing Assistance}
LLMs were used solely to improve the language and clarity of this manuscript. All technical content was written by the authors, and all AI-assisted edits were reviewed and approved by the authors.

\end{document}